\documentclass[aps,twocolumn,epsfig,showpacs]{revtex4-1}
\usepackage{graphicx}
\usepackage{epsfig}
\begin{document}
\title{Scaling of hysteresis loop of interacting polymers under a
periodic force}
\author{Rakesh Kumar Mishra$^{1}$, Garima Mishra$^{1}$, Debaprasad Giri$^{2}$, 
and  Sanjay Kumar$^{1}$ }
\affiliation{$^{1}$Department of Physics, Banaras Hindu University,
     Varanasi 221 005, India \\
$^{2}$Department of Applied Physics, IIT (BHU), Varanasi 221 005, India} 
\begin{abstract}
Using Langevin Dynamics simulations, we study a simple model of  
interacting-polymer under a periodic force. The force-extension curve
strongly depends on the magnitude of the amplitude $(F)$ and the frequency
($\nu$)  of the applied force. In low frequency limit, the system retraces 
the thermodynamic path. At higher frequencies, response time is greater 
than the external time scale for change of force, which restrict the biomolecule 
to explore a smaller region of phase space that results in  hysteresis  of 
different shapes and sizes. We show the existence of dynamical transition, 
where area of hysteresis loop approaches to a  large value from nearly 
zero area with decreasing frequency. The area of hysteresis loop is 
found to scale as $F^{\alpha} \nu^{\beta}$ for the fixed length. 
These exponents  are found to be the same as of the mean
field values for a time dependent hysteretic response to
periodic force in case of the isotropic spin. 
\end{abstract}
\maketitle

\section{introduction}
Many cellular processes are driven by mechanical forces. Synthesis and 
degradation of  proteins \cite{shtilerman, huang}~, transcription and 
replication of nucleic acids, and packing of DNA in a capsid are few examples 
\cite{neuwald, cook, Guo}. In fact, biological motors fueled by 
ATP transform chemical energy to mechanical energy through the hydrolysis 
process \cite{albert,tom}. Periodic consumption
of ATP to ADP suggests that biological motors generate forces of periodic 
nature. For examples, DNA-B, a ring like hexameric helicase that pushes 
through the DNA like a wedge \cite{dnab}. Williams and Jankowsy 
\cite{nphii} showed that viral RNA helicase NPH-II that hops cyclically 
from the double stranded (ds) to the single stranded (ss) part of DNA and
back during the ATP hydrolysis cycle. It is also proposed that pulling force
resulting from ATP consumption is used by proteasomes and mitochondrial to
unfold proteins \cite{wang, Lee, Navon}. 
\begin{figure}[t]
\includegraphics[width=3.0in,clip]{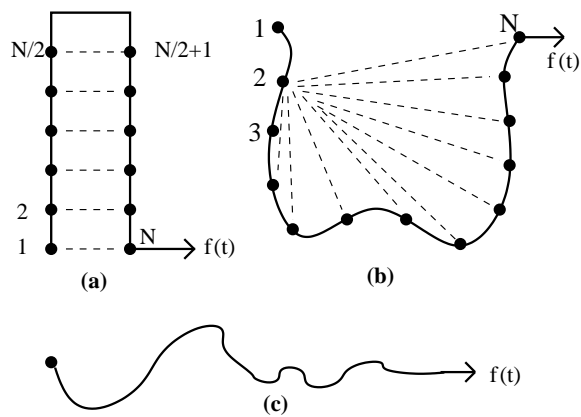}
\caption{Schematic representations of an interacting polymer: 
(a) DNA in the zipped state, (b) a self-interacting polymer chain and 
(c) extended form of an interacting polymer under the influence of
applied force ($f$). In all these cases, one end of polymer is fixed 
and the other end may be  subjected to a constant force or 
periodic stretching force. For DNA,
the dotted lines represent base pairing interaction among complimentary
nucleotides (say 1 to $N/2$ are made up of adenine (A) and $(N/2+1)$ to $N$ are 
made up of complimentary nucleotides i.e. thymine (T)). In this case, 
base pairing interaction is restricted in such a
way that the first monomer forms base pair with the $N^{th}$ monomer and
$2^{nd}$ monomer forms base pair with $(N-1)^{th}$ and so on.
For a self-interacting polymer, the dotted lines show the attractive interaction 
among non-bonded monomers. In this case, any monomer of a chain can interact 
with the rest of monomers of the chain. Here, for example, we have shown the 
second monomer of the chain, which is  interacting with rest of 
non-bonded monomers. Similarly, other monomers interact with rest of 
the non-bonded monomers.
}
\label{fig-1}
\end{figure}

In recent years, there are theoretical efforts to understand the 
response of periodic force on the bio-molecules \cite{Vidybida,Braun,
Pereverzev, Lomholt,Lin}. Most of these studies were 
confined to 
understand the kinetics under the equilibrium conditions \cite{kumarphys}. However, 
living systems are the open systems and never at equilibrium. In the equilibrium, 
bio-molecule follows
the force in phase. The force-extension curve for a periodic force, keeping other 
intensive quantities fixed, would result in retracing thermodynamic path, 
ending at the initial state. In contrast, in the non-equilibrium situation, 
the difference between the relaxation time and the external time scale for
change of force would restrict the bio-molecule to explore a smaller region 
of the phase space, thereby creates
hysteresis i.e. a difference in the response to an increasing and decreasing 
force. Hysteresis is well studied in the context of the spin systems 
\cite{madan1,madan2, dd, bkc}. It was found
that the area of the hysteresis loop scales as $h^{\alpha} \nu^{\beta}$, where
$h$ and $\nu$ are the amplitude and the frequency of the applied magnetic 
field, respectively. The values of $\alpha$ and $\beta$ 
differ from system to system \cite{bkc} and the reason for it has been 
discussed in \cite{siam}. 

Though, hysteresis has been observed in single molecule experiments
\cite{Liphardt,Collin,Schlierf,Hatch,Li1,Born,Friddle,Diezemann}, however,
many aspects of these phenomena are yet to be explored. In a recent
work \cite{Kapri}, Kapri showed that using the work theorem \cite{sadhu}, it is 
possible to extract the equilibrium force-extension curve from the 
hysteresis loops. In another work,  a dynamical transition has been proposed, 
where the area of loop
changes with the frequency of the applied force from nearly zero to a finite 
value, similar to the one seen in case of spin systems \cite{arxiv,arxiv1}. 
In low frequency limit, for a short DNA (16 base pairs), 
the scaling exponents ($\alpha$ and $\beta$) are found to be the same as of 
the isotropic spin system. This raises many questions such as 
does the dynamical transition exist in the thermodynamic limit? Do 
these exponents depend on the length of the DNA? What is the role of molecular 
interaction (involved in the stability of bio-molecules) on the dynamical 
transition.  In this context, it 
is pertinent to mention here that in Ref. \cite{arxiv1}, modeling of DNA 
involves a single polymer chain with native interaction (base pairing) as shown 
in Fig. 1a. Will scaling exponents change, for a self-interacting polymer 
(SIP) chain, where a monomer of the chain can interact (non-native interaction) with the rest 
(Fig. 1b) of monomers of the chain
\cite{kumar_prl}? The present paper addresses some of such issues. In section 
II, we develop a simple  model of polymer and impose certain
constraints to model different bio-polymers. We briefly describe in this section
the  Langevin Dynamics(LD) simulations \cite{Allen,Smith} to obtain the 
thermodynamic observables. In Sec. III, we study equilibrium properties of a short DNA 
and a SIP and obtain the force-extension curves 
and the force-temperature diagrams. Section IV deals with the dynamical 
transition associated with DNA of different lengths. We obtain the 
various exponents associated with the hysteresis loop. We also discuss 
dynamical transition associated with a SIP and obtain 
the scaling exponents. In Sec. V, we discuss finite size scaling.
Sec. VI describes the dynamics near $T =0$, which helped us in identifying
natural frequency to describe the transition. Finally in Sec. VII, we  summarize our 
results and discuss the future perspectives.

\section{Model and Method }

Bio-molecules exhibit a wide range of time scales over which specific 
processes take place \cite{kumarphys}. For example, local motion, which involves atomic 
fluctuation, side chain motion and loop motion, occurs in the length scale 
of 0.01 to 5 \AA{} and the time involved in such a process is of the order 
of $10^{-15}$ to $10^{-12}$s. The motion of  helix and protein domain belong
to the rigid body motion, whose typical length scales are in between 
1 to 100 \AA{} and time involved in such motion is in between $10^{-9}$ to 
$10^{-6}$s. Here, our interest is in the large-scale motion {\it e.g.} 
helix-coil transition, folding-unfolding transition of proteins and 
coil-globule transition in polymer, which occurs in the 
length scale more than 5 \AA{} and time involved is about $10^{-7}$ to $100$ s. 
Since, such a time scale is not amenable computationally, therefore, we consider 
a coarse-grained model of a linear polymer chain and impose restrictive 
interaction among monomers in such a way that it captures some essential 
properties of different bio-polymers \cite{Li,Kouza,MSLi_BJ07,thiru,mishra}.  
We follow Ref. \cite{arxiv}, where the energy of the model system is defined 
by the following expression:
\begin{equation}
E =\sum_{i=1}^{N-1}k(d_{i,i+1}-d_0)^2+\sum_{i=1}^{N-2}\sum_{j>i+1}^{N}
(\frac{B}{d_{i,j}^{12}}-C(\frac{A_{i,j}}{d_{i,j}^6})),
\end{equation}
where, $N$ is the total number of beads/monomers present in the polymer chain. 
The distance between $ i^{th} $ and $ j^{th} $ 
bead is denoted by $d_{i,j}$ $=$  $|\vec r_i-\vec r_j|$, where $\vec r_i$ 
and $\vec r_j$ are the position of bead $i$ and $j$, respectively. 
First term of Eq. 1 is the potential function for covalent bonds between two 
consecutive monomers and is represented by the harmonic potential with   
spring constant $k (=100)$ \cite{mishra}. Second term in the expression 
represents Lennard-Jones (L-J) potential, which models non-bonded interaction 
among monomers of the chain. The first term of L-J potential 
takes care of the excluded volume effect {\it i.e.} two monomers can not
occupy the same space. Second term of the L-J potential gives the  
attractive interaction between all monomers except the adjacent one.
The parameter $d_0 (=1.12)$ corresponds to 
the equilibrium distance in the harmonic potential, which is 
close to the equilibrium position of the average L-J potential.
The values of $A_{ij}$ and $B$ are set to be equal to 1. 
The parameter $C =1.0$ is chosen here for the DNA.
For the SIP, we choose  $d_0 =1.00$, $k = 16.67$ and $C =2.0$, respectively as adopted
in the model discussed in the Ref. \cite{mickal}.

In force induced transitions,  one stretches a bio-polymer
from its ground state (native conformation), therefore, properties associated
with this transition are mainly governed by its native topology. 
The Go Model, which is based on the native-topology is found to be quite useful in
studying the influence of mechanical forces on the bio-polymers \cite{go, Poland, Navin}.
It may be noted that by restricting $A_{ij}$ (second term of L-J potential), 
it is possible to model native topology
of different bio-polymers. For example, if half of a polymer chain
is allowed to interact with the other half of a chain, in such a way that 
the first monomer interacts
only with the $N^{th}$ monomer (last one), and the second monomer interacts with 
the $(N-1)^{th}$ and so, the ground state conformation resembles a zipped conformation
of DNA of $N_p$ ($=N/2$) base pairs as shown in Fig. 1a \cite{arxiv, arxiv1, mishra, Foster}. 
Similarly, if any monomer of the chain is allowed
to interact with the rest of non-bonded monomers of the polymer chain, 
the ground state will resemble the globule (collapsed) state of a 
self-interacting polymer \cite{kumar_prl}. 
The native topology-based model, turns 
out to  be quite helpful in predicting the mechanism involved in the DNA 
unzipping and protein unfolding. It also allowed to decipher the free-energy 
landscapes of bio-polymers\cite{Li,Kouza,MSLi_BJ07,thiru,mishra,pablo}. 
  
The dynamics of the system is obtained from the Langevin equation 
\cite{Allen,Smith,MSLi_BJ07}. It is a stochastic differential equation, 
which introduces friction and noise terms in to Newton's second law to 
approximate the effects of temperature and environment:

\begin{equation}
m\frac{d^2{\bf r}}{dt^2} = -{\zeta}\frac{d{\bf r}}{dt}+{\bf F_c} +\Gamma,
\end{equation}
where $m (=1)$ and $\zeta(=0.4$) are the mass of a bead and the friction 
coefficient, respectively. Here, ${\bf F_c}$ is defined as $-\frac{dE}{d{\bf r}}$ and 
the random force $\Gamma$ is a white noise \cite{Smith}, and is related
to the friction coefficient by fluctuation dissipation theorem
i.e., $<{\Gamma(t)\Gamma(t')}>=2\zeta T\delta(t-t')$. 
It may be noted that the friction term used here only influences the kinetics, 
not the thermodynamic
properties \cite{Li,Kouza}. The choice of this dynamics keeps temperature ($T$) constant 
throughout the simulation. The equation of motion is 
integrated by using the $6^{th}$ order predictor-corrector 
algorithm with time step $\delta t$=0.025 \cite{Smith} for DNA and $\delta t$=0.005
for the SIP, respectively. The results are averaged over many trajectories. 

\section{Equilibrium properties of bio-polymers}

\begin{figure}[t]
\centerline{\epsfig{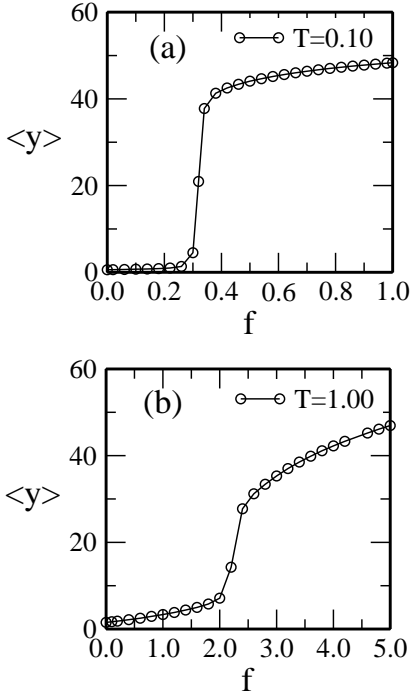}}
\caption{(a) Equilibrium force-extension ($f-y$) \\ 
curve for DNA and (b) for a self-interacting polymer.
}
\label{fig-2}
\vspace {0.5cm}
\end{figure}
The equilibrium properties of DNA unzipping has been studied in
the constant force ensemble (CFE) for the fixed length \cite{mishra}. Here,
we also study the equilibrium properties of 
self-interacting polymer in CFE and compare its result with DNA.
The equilibrium has been checked by calculating  the auto-correlation 
function \cite{box,berg} for the  square of the end-to-end distance and the 
number of contacts \cite{mishra}. Moreover, these results are tested over 
many seeds and the 
stability of data against at least ten times longer run. We have 
used $2\times10^9$ time steps out of which first $ 5\times 10^8$ steps are 
not taken in the averaging. In the equilibrium condition, we add an energy 
$-\vec{f}.\vec{y}$ to the total energy of the system given by Eq. 1 
\cite{mishra}. 
We calculate the average  extension $<y>$ (distance between the end  monomers) 
at different values of $f$. The force-extension curves (Fig. 2) for
both cases show qualitatively similar behavior.  Below the critical
force, the systems remain in the zipped (or globule) state, and above it,
in the unzipped (or coil) state. It may be noted that, if we decrease 
the force keeping the other intensive parameters fixed, the system nearly 
retraces the path showing that it is in the equilibrium.

The equilibrium force-temperature ($f-T$) diagram may be obtained by 
monitoring the energy fluctuation (or the specific heat) with  
temperature at different forces $f$ \cite{Li,Kouza,mishra}. The peak in the 
specific heat curve gives 
the melting temperature at that $f$, which is consistent with  
the $f-y$ curve for that temperature $T$. The phase boundary in the $f-T$ diagrams 
(Fig. 3) separates the region, where the DNA (or SIP) exists in a 
zipped (or globule) state from the region, where it exists in the unzipped 
(or coil) state.
It is evident from these plots (Fig. 3) that the melting temperature 
decreases with the applied force in accordance with the earlier studies 
\cite{bhat99,marenprl}. We find that the peak height 
increases with the chain length, though, the transition temperature 
(melting temperature) remains almost the same for different lengths.

\begin{figure}[t]
\hspace{1.0 cm}\centerline{\epsfig{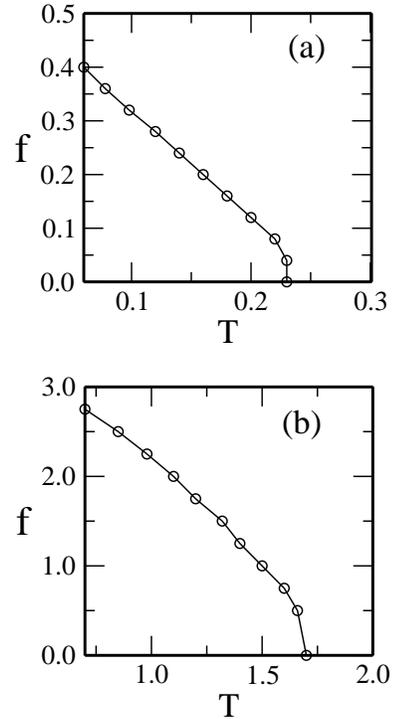}}
\caption{(a) Equilibrium force-temperature ($f-T$) diagram of DNA 
and (b) for a self-interacting polymer.
}
\label{fig-3}
\vspace {0.5cm}
\end{figure}

\begin{figure*}[ht]
\includegraphics[width=4.5in,clip]{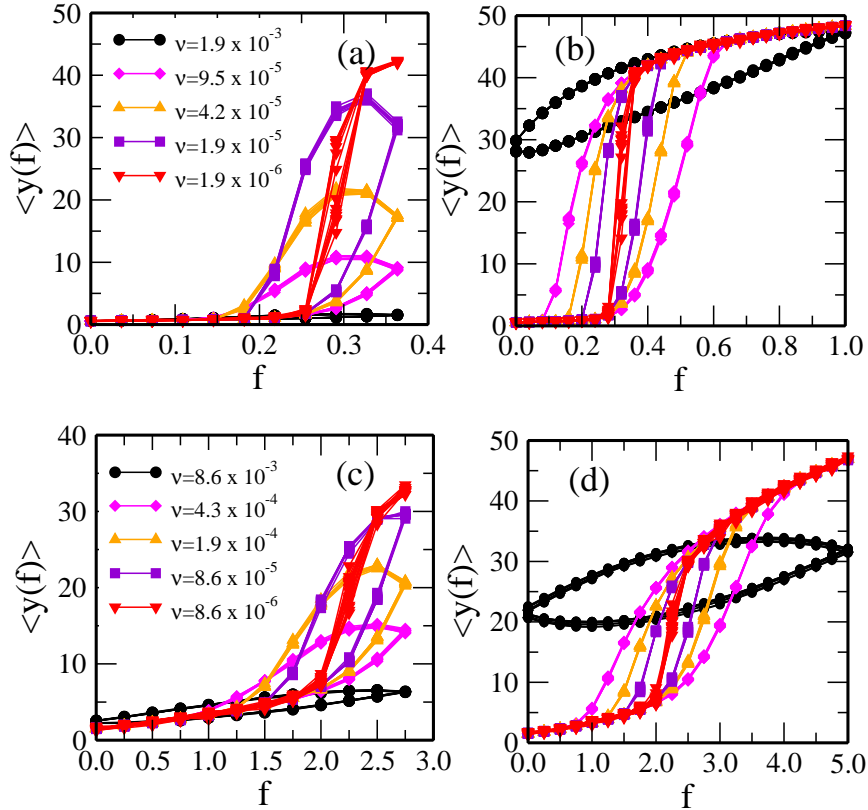}
\caption{The averaged extension of DNA as a function of 
cyclic force of amplitude (a) 0.4 and (b) 1.0 at different $\nu$.
Figs. (c) and (d)  are for the SIP for  low (2.75) and high amplitude (5.0),
respectively. It is evident from these plots (Figs. a \& c)
that at low amplitude and high frequency, the system remains in the zipped 
(collapsed) state, whereas at high amplitude
and high frequency (Figs. b \& d), the system remains in the extended state with a
small hysteresis loop. As $\nu$ decreases, both the systems extend to the 
hysteretic state with a bigger loop. For $\nu \rightarrow 0$, the hysteresis
loop vanishes and the system approaches its equilibrium path irrespective
of the magnitude of amplitudes of the applied force.
}
\label{fig-4}
\end{figure*}

\section{Dynamical transition at finite temperature}
In order to study the dynamical stability of the bio-molecule under a 
periodic force \cite{arxiv}, we add an energy $-f(t).y(t)$ to the total energy
of the system given by Eq. 1.  The value of $f$ increases to its maximum value $F$ 
in  $m_s$ steps at  interval  $\Delta f (= 0.01)$ and then 
decreases to $0$ in the same way \cite{arxiv,arxiv1}. Since, we are interested 
in the non-equilibrium regime, we allow only  $n$  LD time steps 
(much below the equilibrium time) in each increment of $\Delta f$. 
Here, $y(t)$ is the distance between the two ends of the chain at that instant
of time. We keep sum of the time spent $\tau (= 2nm_s)$ in  each force cycle 
constant to keep $\nu (=1/\tau)$ constant. Hence, for a given $F$,
$n$ controls the frequency. By varying $F$ (keeping $\nu$ fixed) or $\nu$
(keeping $F$ fixed), it is possible to induce a dynamical transition between
a time averaged zipped (globule) state or unzipped (coil) state
to a hysteretic state (oscillating between the zipped and the unzipped state).

The average extension of the  DNA (Fig. 4 a \& b) and SIP (Fig. 4 c \& d)
clearly exhibits hysteresis under the periodic force. We have performed
average over 1000 cycles for different initial conformations. In Figs,(4a-d) we
have shown the plots for 10 different initial conformations. All these
curves overlap indicating that the system is in the steady state, irrespective
of the starting conformation (zipped or unzipped). 
If  the time averaged $y(f)$ is less than $5$, we call
the system is in the zipped state, where as if $y(f) > 5$, it
is in the open state or in the unzipped state \cite{arxiv,arxiv1}.
At low amplitude, the chain remains in the zipped state (Fig. 4a), with almost
negligible loop area. As the frequency decreases, the system
explores more phase space and acquires conformations belonging to the 
unzipped state. We calculate the dynamical order parameter \cite{bkc} 
i.e. the area
of the hysteresis loop, which is defined as
\begin{equation}
A_{loop}= \oint y(f).df.
\end{equation}

One may notice from Fig. 4 
that the area of loop increases as the frequency decreases. After a certain 
frequency, the area of the loop starts decreasing, and the system nearly 
retraces 
the equilibrium path at low frequency. The self-interacting polymer, 
which shows the existence of globule state (Fig. 2) at low $T$, exhibits 
the similar behavior (Fig. 4 c \& d). However, at high amplitude, where 
the dsDNA shows the existence of the stretched state for all frequencies, SIP shows 
the existence of two states i.e. extended and stretched. Since, for the 
SIP the ground state energy is quite large compare to DNA, the 
unfolding force is also found to be larger than the dsDNA. It is 
in accordance with the experiments \cite{kumarphys}.
The other interesting observation from these plots is that
though $f$ decreases from its maximum value F to 0 (Fig. 4 a \& c),
$y(f)$ increases and there is some lag, after which it decreases.
One may recall that the relaxation time is much higher compare to the time
spent at each interval of $\Delta f$, and therefore, an increase in $y(f)$ with
decreasing $f$, indicates that the system gets more time to relax. As
a result $y(f)$ approaches a path, which is close to the equilibrium.
Once the system gets enough time, the lag disappears. A similar lag can be 
seen, when the system starts from the open state at high $\nu$.
However, in this case as $\nu$ decreases, $y(f)$ decreases with increasing $f$.
In both the cases , whether DNA starts from the zipped or open state,
as  $\nu \rightarrow 0$, the system approaches the equilibrium
$f-y$ curve and the area of loop vanishes.

\begin{figure}[t]
\includegraphics[width=2.2in,clip]{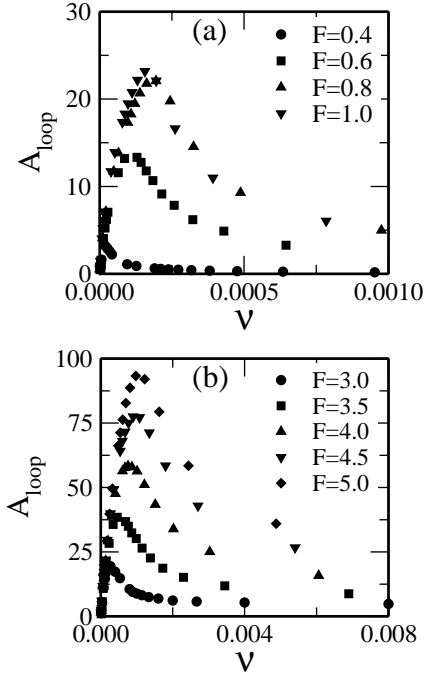}
\caption{Figures show the variation of area of hysteresis loop with the frequency 
at different force amplitudes (a) for DNA and (b) for SIP. For both cases,
the area of loop increases to its maximum with the frequency and then 
again approaches to zero. In these cases, the system approaches the equilibrium
from the non-equilibrium as the frequency decreases.
}
\label{fig-5}
\end{figure}

In Fig. 5, we have plotted the area of hysteresis loop of DNA (Fig. 5a) and
SIP (Fig. 5b) with 
the frequency at different force amplitudes. One can see from these plots 
that the area of hysteresis loop increases with decrease in the frequency 
and starts decreasing after a certain frequency. At a very low frequency, 
the system approaches its equilibrium i.e. the extension  nearly retraces its path 
under the cyclic force.  Fig. 6 a \& b shows the variation of loop area for 
DNA and SIP, respectively, with force amplitude at different frequencies. 
Here, the area also increases with the amplitude and after
certain amplitude, it starts decreasing similar to  Fig. 5 . In this case, 
however, the system goes far away from the equilibrium. 

\begin{figure}[t]
\includegraphics[width=2.2in,clip]{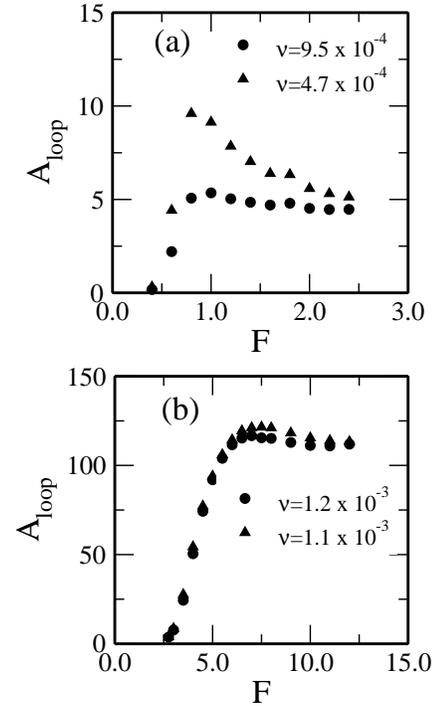}
\caption{Variation of the area of hysteresis loop with force amplitude at different
frequencies: (a) for DNA and (b) SIP. In this case, the system never approaches 
to equilibrium and always remains far away from the equilibrium as the amplitude of 
force increases.
}
\label{fig-6}
\end{figure}

\begin{figure}[t]
\includegraphics[width=3.2in,clip]{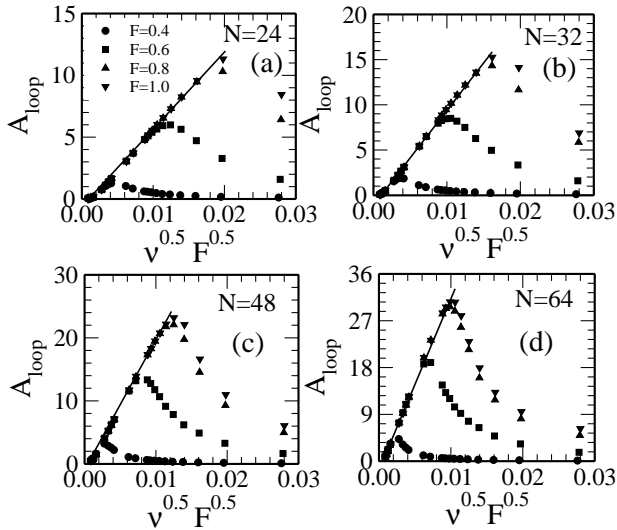}
\caption{ Figs (a-d) show the scaling of loop area of hysteresis with 
respect to  $\nu^{0.5}F^{0.5}$ for different lengths. It is evident from 
all these plots that in low frequency limit curves of different amplitudes 
collapse on a single line intricating that the dynamical transition may exist 
in the thermodynamic limit.  
}
\label{fig-7}
\end{figure}

For the spin systems, the area under the hysteresis loop ($ A_{loop} 
\sim h^\alpha \nu^\beta$) is the measure of energy dissipated over a 
cycle. 
In a recent paper, Kumar and Mishra \cite{arxiv1} for a small dsDNA ($N_{p}=16$ base pairs) 
found a similar scaling for DNA unzipping i.e. the area of hysteresis 
loop scales with $F^{\alpha}\nu^{\beta}$. In low frequency regime, 
value of $\alpha$ and 
${\beta}$ are found to be equal to $0.5$, which are same as one obtained 
in the case of isotropic spin system \cite{dd}. At high frequency, 
the values of $\alpha$ and $\beta$ 
are found to be $2$ and $-1$, respectively, which are also consistent with
isotropic spin system. 
In order to see whether these scaling holds for different lengths, 
we measured the area of hysteresis loop for the various
chain length of DNA ($N=24, 32, 48$ and $64$)  and plotted it with $F^{0.5}\nu^{0.5}$ 
in the low frequency regime (Fig. 7 a-d) and  $(F-f_{c})^{2.0\pm0.1}\nu^{-1}$ in
 the high frequency regime (Fig. 8 a-d). Here, $f_{c}$ is the equilibrium critical 
force at that temperature (Fig. 3). For all lengths studied here, the collapse 
of data for different $F$ on a single curve in low and high frequency regimes, 
suggest that the dynamical transition may be seen in single molecule experiments,
which involve chain of finite size. 
\begin{figure}[b]
\includegraphics[width=3.2in,clip]{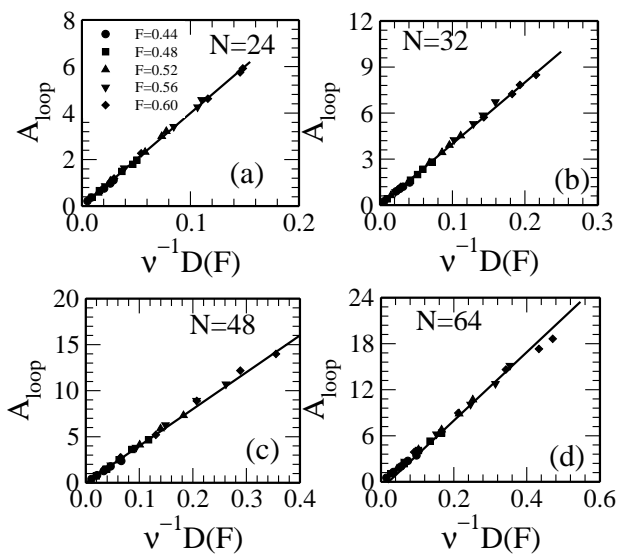}
\caption{Same as Fig.7, but loop area of hysteresis has been plotted
with respect to $\nu^{-1.0}D(F)$. Here D(F) $\sim$ $(F-f_{c})^{2.0\pm0.1}$. 
In high frequency regime also, curves
of different length of DNA collapse on a single line.
}
\label{fig-8}
\end{figure}

We now focus our study to SIP under a periodic force, where a bead (or monomer) 
can interact with the rest of non-bonded monomers. It is interesting to note that
in low as well as in high frequency regimes, the SIP also obey the similar
scaling  as shown in the Fig. 9 a \& b with the same scaling exponents.

\begin{figure}[t]
\includegraphics[width=2.4in,clip]{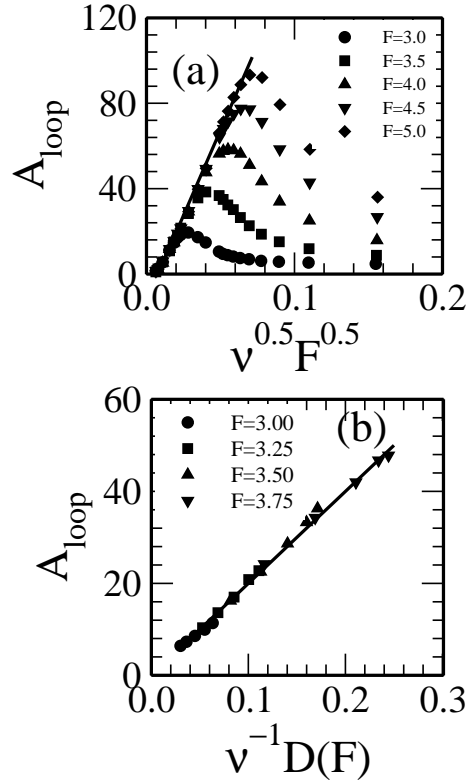}
\caption{(a) Same as of Fig. 7, but for the SIP. In low frequency regime, curves
of different amplitudes collapse on a single line.
(b) At high frequency, curves for different F collapse on a straight line. Here
D(F) $\sim$ $(F-f_{c})^{2.0\pm0.1}$. 
}
\label{fig-9}
\end{figure}

\section {Finite size scaling} 

We have studied the average energy dissipated per monomer i.e. $A_{loop}/N$, over
a cycle. In the limit $\nu \rightarrow 0$, we have plotted the  $A_{loop}/N$ with  
$F^{0.5}\nu^{0.5} N^{0.75}$ in Fig. 10. One can see the collapse of data for all lengths
of different forces and frequencies. Similarly, Fig. 11 shows the collapse of data
of all lengths in high frequency regime. It is interesting to note that in high 
frequency regime the average dissipated energy ($A_{loop}$) is independent of length. 
This may be understood by realizing that the applied force will try to move the end 
terminal with a velocity $v$. For a very short duration of time ($\nu \rightarrow 
\infty$), the applied force can move the end terminal to a finite distance, which 
is independent of length. Hence, the area under a cycle of force 
(0 to $F$ and back to 0) will remain independent of length (Fig. 11). However, 
in low frequency regime ($\nu \rightarrow 0$), the system gets enough time 
(close to equilibrium). Because of the connectivity of the beads in the chain, 
the applied force will be transmitted all along the chain. Consequently, 
both strands will move under the  periodic force and the resultant 
curve under hysteresis will depend on the length of the chain (Fig. 10).

\begin{figure}[ht]
\includegraphics[width=2.4in,clip]{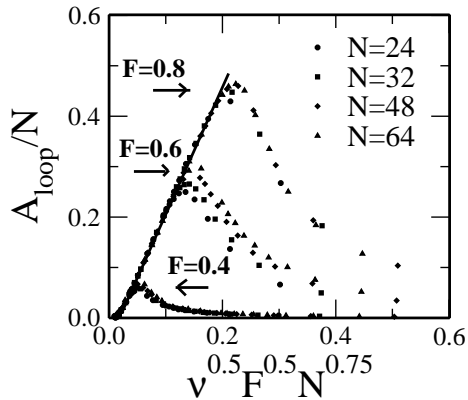}
\caption{Figure shows the collapse of data for different lengths. In low frequency 
regime, curves of different amplitudes collapse on a single line.
}
\label{fig-10}
\end{figure}

\begin{figure}[h]
\includegraphics[width=2.4in,clip]{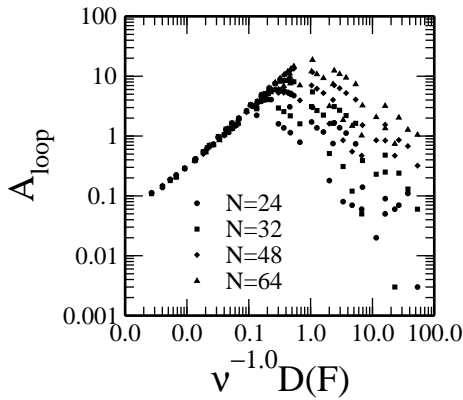}
\caption{Figure shows that at high frequency, curves for different lengths 
collapse on a straight line. (For clarity we have plotted Fig.10 
in log-log scale)
Here, $D(F)$ has the same scaling form as mentioned in caption of Fig. 8.
}
\label{fig-11}
\end{figure}

\section{Critical frequency and its dependence on $N$}

From Fig. 5, it is clear that as frequency decreases, the  system goes from one 
regime ($\nu^{-1}$) to the other ($\nu^{0.5}$). The frequency at which the area of 
hysteresis loop is maximum (discontinued) is termed as critical frequency 
$\nu_c$ \cite{dd}. 
In Fig. 12 a, we have shown that the critical frequency decreases with N
for a given value of $F$. In order to see its dependence on $N$, we 
investigate the dynamics of the system  near $T = 0$ under a linear ramp (triangular 
shape), similar to the one taken in our simulation \cite{arxiv1}. 
We assume here that there is no acceleration and beads move with a 
uniform velocity (because of environment) under the applied force. 
The contribution of random noise is negligible and Eq. 2 reduces to

\begin{equation}
\zeta\frac{d{\bf r}}{dt} = {\bf F_c} + {\bf F} \nu t.
\end{equation}
The frequency at which maximum area occurs, the extension
$r$ scales as $N$. Since, $F$ is constant therefore $t$ also scales as $N$ and
$\nu$ scales as $1/N$ to keep $\nu t$ constant. In Fig. 12 b, we have plotted
$A/N$ with the scaled frequency $\nu^* =  N \nu$. It is evident from 
Fig. 12 b that maxima of all lengths occur at critical frequency $\nu_c$ 
without changing the qualitative features of the curve. In low frequency 
regime, $A/N$ scales $\nu^{*0.5}$. One can see from the inset of Fig. 12c
that the additional multiplication of $N^{0.25}$ with  $\nu^{*0.5}$ gives a better
collapse similar to the one shown in Fig.10. This may be a crossover or 
effect of random fluctuation because of finite temperature or the entropy 
associated with polymer chain, or the combined effect of all these, whose  
precise contributions  have been ignored in Eq. 4.

\begin{figure}[ht]
\includegraphics[height=5.5in,clip]{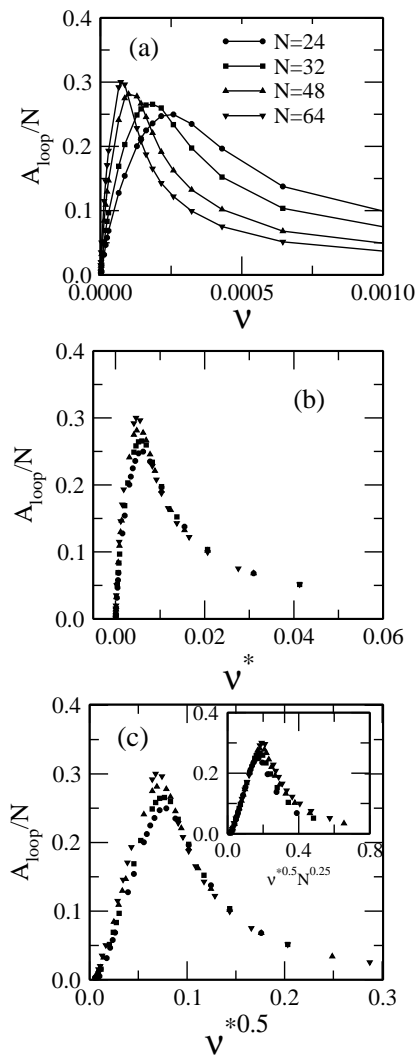}
\caption{(a) Area of the loop per bead with frequency at fixed amplitude
of force $0.6$. The peak of the curve shifts right with $N$. (b) Same as
(a) but with scaled frequency $\nu^* = N \nu$. The peak of all the curves
collapsed at frequency $\nu_c$. (c) In low frequency regime, $A/N$ scales as
$\nu^{*0.5}$. The inset shows  the scaling ($\nu^{*0.5}N^{0.25}$) of x-axis
 gives a better collapse.
}
\label{fig-12}
\end{figure}

\section{conclusions} 
In this paper, we have investigated the dynamical transition associated with
DNA and SIP. A simple model of a polymer developed
here can describe some of the essential properties of bio-polymers
(e.g. DNA and SIP)  depending upon the interaction imposed on the
non-bonded monomers. We have studied these two models under
a periodic force and showed that they show similar behavior. 
Our studies provide strong evidence that
in the low  as well as in the high frequency
regime, the area of hysteresis loop scales with the same
exponents irrespective where a monomer interacts with native neighbor (DNA) or 
the non-native neighbor (SIP). The values of $\alpha$ and $\beta$ are 
consistent with the previous studies and have the same values as found
in case of the isotropic spin systems \cite{dd}. We have also shown that the value of 
these 
exponents remain the same for different lengths of DNA. In high frequency regime, 
area of the loop remains independent of length i.e. $\sim F^{2} \nu^{-1}$. 
In low frequency regime, we report a new scaling, where average energy 
dissipated per bead over a cycle scales as $F^{0.5} \nu^{0.5} N^{0.75}$. 
This suggests that the dynamical transition may be seen in single molecule 
experiments of finite chain. 
However, the present scaling is at finite temperature, where noise in Eq. 2
has an important role. It would be interesting to probe these scaling 
for a longer chain in the low temperature regime, where noise has no significant 
contribution. Our work calls for further investigations whether these exponents are 
universal for the other polymeric models or its value may differ e.g. for proteins 
and RNA, which have distinct structures.

\section{acknowledgments}

We thank  S. M. Bhattacharjee for many helpful discussions 
on the subject. Financial supports from the DST and CSIR, India are gratefully 
acknowledged. We also acknowledge the generous computer support from IUAC, New Delhi.

\end{document}